\documentclass[twocolumn,aps,showpacs,prb]{revtex4-1}

\usepackage{graphicx}
\usepackage{amsmath}
\usepackage{multirow}

\begin{document}

\title{Pressure-induced topological phase transition in LaSb: First-principles study}

\author{Peng-Jie Guo$^{1,2}$}
\author{Huan-Cheng Yang$^{1,2}$}
\author{Kai Liu$^{1,2}$}\email{kliu@ruc.edu.cn}
\author{Zhong-Yi Lu$^{1,2}$}\email{zlu@ruc.edu.cn}

\affiliation{$^{1}$Department of Physics, Renmin University of China, Beijing 100872, China}
\affiliation{$^{2}$Beijing Key Laboratory of Opto-electronic Functional Materials $\&$ Micro-nano Devices, Beijing 100872, China}

\date{\today}

\begin{abstract}
 By using first-principles electronic structure calculations, we predict that the extreme magnetoresistance (XMR) material LaSb takes a topological phase transition without breaking any symmetry under a hydrostatic pressure applied between 3 and 4 GPa, meanwhile the electron-hole compensation remains in its electronic band structure. Thus LaSb provides an ideal platform for studying the individual role of topological property playing in the XMR phenomenon, in addition to the electron-hole compensation. This has general implication to the relationship of XMR effect and topological property in topological materials.
\end{abstract}

\date{\today} \maketitle

\section{INTRODUCTION}

The topological properties of realistic materials have attracted intensive attention both theoretically and experimentally in recent years \cite{Hasan, Qi}. With protection of time-reversal symmetry, insulators can be classified into conventional band insulators and topological insulators \cite{Hasan, Qi}. Compared with conventional band insulators, topological insulators have robust conducting surface states against local perturbations \cite{Hasan, Qi}. Conceptually, a natural generalization from insulator to metal divides semimetals into trivial and nontrivial semimetals in topological electronic structure, with the latter including Weyl, Dirac, and node-line semimetals \cite{Hong}. On the other hand, different topologies are closely related \cite{Hong}. For instance, Dirac semimetal can be transformed into Weyl semimetal by breaking time-reversal or space inversion symmetry \cite{Hong}. More interestingly, the phase transition between topological insulator and conventional band insulator does not break any symmetry, which surpasses Landau$^{'}$s phase transition theory \cite{Thouless, Wen}. This is an important breakthrough in physical theory.

One striking property of topological materials is its magnetoresistance. For topological insulators, the gapless surface states possess many novel properties: their unique spin textures forbid the electron$^{'}$s backscattering under zero magnetic field \cite{Hasan}. For Weyl and Dirac semimetals \cite{HongNa3Bi, ChenS, HongCdAs, ChenN, HongPRX, DingPRX, HasanS, DingNP}, their fermions result in high carrier mobility as well as linear magnetoresistance \cite{Abrikosov, Neupane, LiangNM, YanNP}, meanwhile their chiral anomalies render negative magnetoresistance \cite{Nielsen, VishwanathPRX, XiongS, HuangPRX}. Recently, the extreme magnetoresistance (XMR) as a quadratic function of magnetic field has been discovered in both topologically trivial material LaSb \cite{CavaNP, LouPRL} and its isostructural topologically nontrivial counterpart LaBi \cite{LeiNJP}. Previously, the electron-hole compensation \cite{CavaN} and  backscattering-forbidden topological protection mechanisms \cite{FengPRL} have been proposed to explain the XMR effects, but a consensus has not yet been achieved\cite{ShenPRL}. Through comparative studies on LaSb and LaBi, it has been realized that the electron-hole compensation plays an important role in the XMR \cite{LouPRL, GuoPRB}, but the role of topological property is still unclear \cite{FengPRB}. Searching for a material, which shows the XMR effect with a topological phase transition under certain experimental condition, is thus very important for understanding the relationship of XMR effect and topological property.

Except for topological property, LaSb and LaBi are very similar in both crystal structure and electronic band dispersion \cite{GuoPRB}. These similarities lead us to consider that LaSb may possess the same topological property as LaBi by changing lattice constants or via chemical doping. Nevertheless, chemical doping may ruin the compensation between electron-type and hole-type carriers, which is disadvantageous for studying the relationship of XMR effect and topological property. Instead, external pressure is a clean and powerful approach for tuning electronic structures and studying novel physical properties \cite{GaoPRB, ChiPRL}.

In this article, we have studied the evolution of electronic structure of the XMR material LaSb with hydrostatic pressure by using first-principles electronic structure calculations. We predict that LaSb takes a topological phase transition without breaking any symmetry under 3$\sim$4 GPa. Our calculations indicate that the band inversions at the X points in Brillouin zone result in the nontrivial topological properties of LaSb under pressure.

\section{COMPUTATIONAL DETAILS}
\label{sec:Method}

To study the evolution of electronic structure of LaSb under pressure, we carried out first-principles electronic structure calculations with the projector augmented wave (PAW) method \cite{ Blochl, Kresse1999} as implemented in the VASP package \cite{ Kresse1993, KresseCMS, Kresse1996}. For the exchange correlation functional, we adopted two different levels in Jacob$^{'}$s ladder \cite{PerdewJPC}: the generalized gradient approximation (GGA) and the hyper-GGA, in which the Perdew-Burke-Ernzerhof (PBE) type of formulas \cite{PerdewPRL} and the screened hybrid functional \cite{hybrid} introduced by Heyd, Scuseria, and Ernzerhof (HSE) with the HSE06 version \cite{hse06} were used, respectively. The kinetic-energy cutoff of the plane-wave basis was set to be 300 eV. For the Brillouin-zone (BZ) sampling, an 8$\times$8$\times$8 $k$-point mesh was adopted for the primitive cell, which contains one formula unit, of the rock-salt structural crystal. The Fermi level was broadened by a Gaussian smearing method with a width of 0.05 eV. Both cell parameters and internal atomic positions were fully relaxed. The system under hydrostatic pressures in a range of 0 to 14 GPa was simulated by assigning the converged trace of stress tensor to a targeting pressure and minimizing the enthalpy of the system \cite{Ye13PRB}. The atoms were allowed to relax until all forces were smaller than 0.01 eV/\AA. After the equilibrium crystal structures at different pressures were obtained, the electronic structures were studied by including the spin-orbit-coupling (SOC) effect. The topological invariants were calculated using the parities of all filled states at eight time-reversal-invariant points in BZ \cite{Fu07PRB}.

\section{RESULTS AND ANALYSIS}
\label{sec:Results}

To study the evolution of electronic structure of LaSb with pressure, we first need to clarify its crystal structures under different pressures. Previous experiments have found that LaSb undergoes a structural phase transition from a NaCl-type lattice [Fig. 1(a)] to a body-centered tetragonal lattice [Fig. 1(b)] around 11 GPa \cite{Leger}. We have thus calculated the pressure-dependent enthalpies of LaSb in these two forms of crystal structures [Fig. 1(c)]. At ambient pressure, the enthalpy of the rock-salt structure is lower than that of the body-centered tetragonal structure, suggesting that the former is more stable. With increasing pressure, the enthalpies of both structures rise up gradually. However, the enthalpy of the rock-salt structure grows faster than that of the body-centered tetragonal structure, and around 9 GPa there exists a crossover. This result verifies that LaSb takes a pressure-induced structural phase transition at $\sim$9 GPa, which is in accord with previous measurements \cite{Leger}.

\begin{figure}[!t]
\includegraphics[width=1\columnwidth]{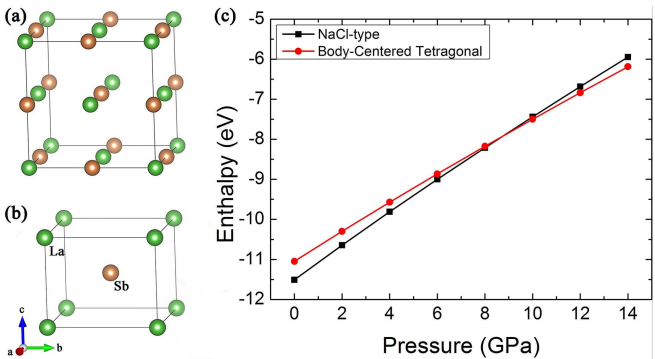}
\caption{(Color online) (a) The NaCl-type and (b) the body-centered tetragonal crystal structures of LaSb. (c) Enthalpies of LaSb in these two forms of structures as a function of pressure.}
\label{Fig1}
\end{figure}

\begin{figure}[!t]
\includegraphics[width=0.9\columnwidth]{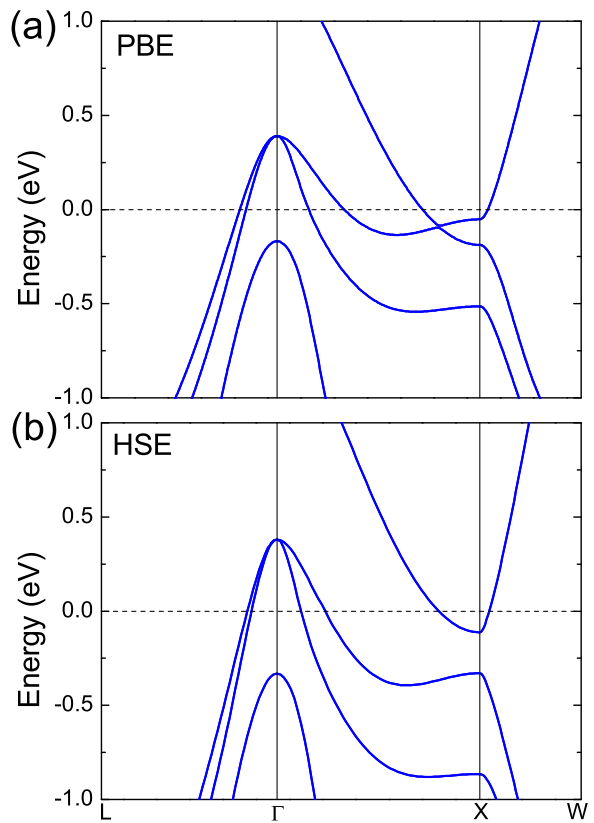}
\caption{(Color online) Band structures of LaSb in NaCl-type structure along high-symmetry directions of Brillouin zone calculated with the (a) PBE and (b) HSE06 functionals including the spin-orbit-coupling (SOC) effect at ambient pressure. The Fermi level is set to zero.}
\label{Fig2}
\end{figure}

Once the crystal structure is determined, the band structure of LaSb can be studied under pressure. We first focus on the rock-salt structure at ambient pressure. It is well known that La and Sb atoms have three and five valence electrons, respectively. If Sb atom can completely obtain three valence electrons from La atom, LaSb should be an insulator. We have investigated the band structure of LaSb along high-symmetry directions in the BZ by using the PBE and HSE06 functionals and including the SOC effect (Fig. 2). Here, only the bands around Fermi level are presented. We find that the valence and conduction bands of LaSb have a small overlap, demonstrating that LaSb is a semimetal, which is in agreement with previous transport experiment \cite{CavaNP}. On the other hand, there is a band inversion around the high-symmetry point X in BZ for the band structure calculated with the PBE functional at GGA level [Fig. 2(a)], but no band inversion with the HSE06 functional at hyper-GGA level [Fig. 2(b)]. Our previous calculations with the modified Becke-Johnson (MBJ) exchange potential \cite{Becke06, Tran09} at meta-GGA level also indicate no band inversion around the X points \cite{GuoPRB}. Since the band inversion is a very important feature for nontrivial topology, LaSb shows different topological properties in calculations at different exchange-correlation functional levels. From the experimental side, the angle-resolved photoemission spectroscopy (ARPES) measurement clearly demonstrates that there is no band inversion around the X points, showing that LaSb is a topologically trivial semimetal at ambient pressure \cite{LouPRL}. This demonstrates that  the calculated band structures at higher (meta-GGA and hyper-GGA) levels agree better with the one obtained in ARPES experiment.

We know that different from chemical doping, pressure can effectively tune the electronic structures of materials without introducing foreign atomic species. So one may wonder whether the nontrivial topology would tale place in the electronic structures of LaSb with increasing pressure, especially before the structural phase transition. In order to answer this question, we have calculated the band structures of LaSb under 1 GPa to 8 GPa, in which the rock-salt structure and the HSE functional were used as suggested by above results (Figs. 1 and 2). We find that from 1 to 3 GPa there is no band inversion around  the X points, but from 4 to 8 GPa the band inversion takes place. For simplicity, only the representative band structure at 3 GPa before the transition [Fig. 3(a)] and that at 4 GPa after the transition [Fig. 3(b)] are displayed.

\begin{figure}[!t]
\includegraphics[width=0.9\columnwidth]{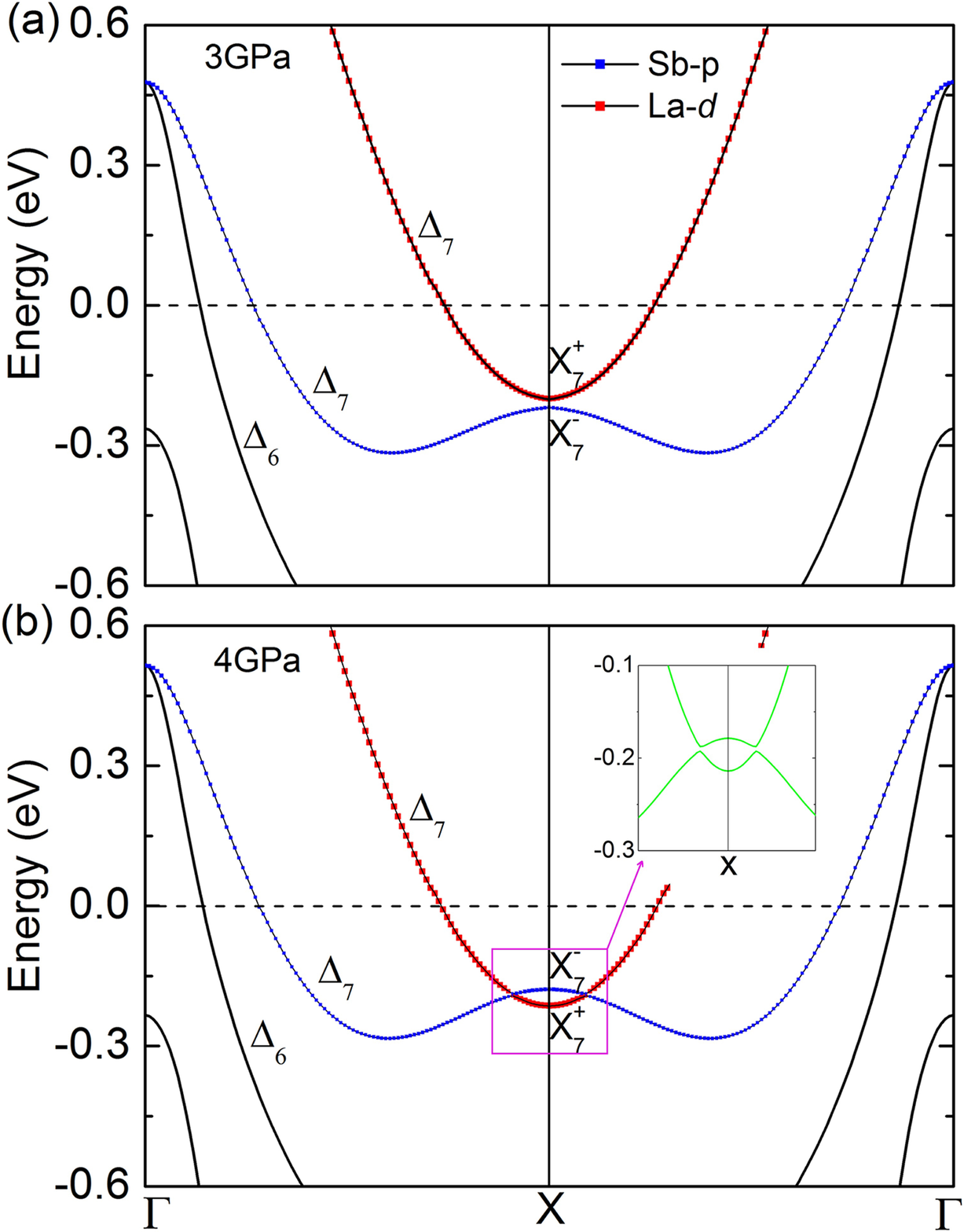}
\caption{(Color online) Orbital characteristics and symmetries of the two bands around the X points near the Fermi level for LaSb in NaCl-type structure calculated with the HSE06 functional and including the SOC effect at (a) 3 GPa and (b) 4 GPa. The Fermi level is set to zero. The signs $\pm$ indicate the parities of corresponding states. Inset shows the enlarged region around the band crossing.}
\label{Fig3}
\end{figure}

More information is needed to get in-depth insight into these two (crossing) bands around the X points (Fig. 3). The symmetry for the $\Gamma$-X direction in BZ of LaSb in rock-salt structure is C$_{4v}$ double group when the SOC effect is included. In combination with the fact that LaSb has both time-reversal and space-inversion symmetries, the band crossing around the X points may render LaSb a Dirac semimetal. Accordingly, we analyze the related orbital weights, symmetries, and wave function parities of these two crossing bands. With orbital weight analysis of the bands near the Fermi level along the $\Gamma$-X direction, the valence (lower) band mainly comes from Sb $p$ orbitals, while the conduction (upper) band is primarily contributed by La $d$ (t$_{2g}$) orbitals. Before the band crossing at 3 GPa [Fig. 3(a)], the symmetries of the valence band are $\bigtriangleup_{7}$ along the $\Gamma$-X direction and X$^{-}_{7}$ (odd parity) at the X points respectively, while those of the conduction band are $\bigtriangleup_{7}$ along $\Gamma$-X direction and X$^{+}_{7}$ (even parity) at the X points respectively. Since the symmetries of the valence and conduction bands belong to two equivalent irreducible representations of the C$_{4v}$ double group, their crossing along the $\Gamma$-X direction under 4 GPa [Fig. 3(b)] opens a gap when the SOC effect is included (see the inset). This indicates that LaSb is not a Dirac semimetal. However, as there are inversed bands with opposite parities around three equivalent X points in the BZ under 4 GPa [Fig. 3(b)], LaSb may be a topological insulator defined on a curved Fermi surface.

\begin{figure}[!t]
\includegraphics[width=0.9\columnwidth]{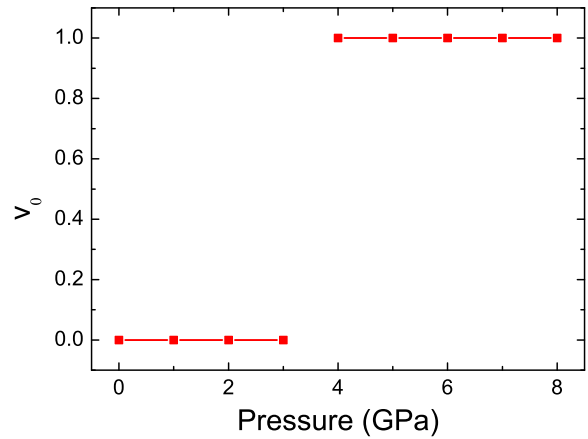}
\caption{(Color online) The topological invariants $\nu_{0}$ of LaSb in NaCl-type structure as a function of pressure.}
\label{Fig4}
\end{figure}

In order to verify the topological property of LaSb under pressure, we have further calculated its Z$_{2}$ topological invariants. Different from two-dimensional quantum spin hall effect (QSHE), three-dimensional topological insulator has four Z$_{2}$ topological invariants \cite{FuPRL}. If the materials own both time-reversal and space-inversion symmetries, the Z$_{2}$ topological invariants can be calculated from the parities of filled states at all time-reversal-invariant points. The value (1 or 0) of the first Z$_{2}$ topological invariant, $\nu_{0}$, indicates whether a topological insulator is strong or weak. Here, we have calculated this important index $\nu_{0}$ as a function of pressure as displayed in Fig. 4. As we see around the X points, there is no the band inversion at 3 GPa and the $\nu_{0}$ is 0, while at 4 GPa the band inversion takes place and  the $\nu_{0}$ changes to 1. All detailed parities of the filled states at eight time-reversal-invariant points before and after the jump of $\nu_{0}$ are listed in Tables I and II, respectively. By comparing the parities of all these states, we find that the inversion of two bands around the X points nearest to the Fermi level, as showed in Fig. 3, leads to the emergence of topological phase transition. These results demonstrate that there is a topological phase transition before structural transition in NaCl-type LaSb with increasing pressure.

\section{Discussion}
\label{sec:discussion}

\begin{table}[!t]
\caption{\label{tab:I} The parities of all filled states of LaSb in NaCl-type structure at eight time-reversal-invariant points in the Brillouin zone before the topological phase transition.}
\begin{center}
\begin{tabular*}{0.99\columnwidth}{@{\extracolsep{\fill}}cccccccccc}
\hline\hline
  &      $\Gamma$ & L & L & L & L & X & X & X & total  \\
\hline
  1 & - & - & - & - & - & - & - & - & + \\
  3 & - & - & - & - & - & - & - & - & + \\
  5 & - & - & - & - & - & - & - & - & +  \\
  7 & + & - & - & - & - & + & + & + & + \\
  9 & - & + & + & + & + & - & - & - & + \\
  11 & - & + & + & + & + & - & - & - & +  \\
  13 & - & + & + & + & + & - & - & - & + \\
  total & + & + & + & + & + & + & + & + & 0 \\
\hline\hline
\end{tabular*}
\end{center}
\end{table}

Previous studies have shown that LaSb and LaBi possess different topological properties at ambient pressure \cite{LouPRL, GuoPRB, NayakNC, LouPRB}. Given that LaSb and LaBi own the same crystal structures and valence electrons, one may straightforwardly  think that they should give out similar band structures. Nevertheless, a distinct difference between the band structures of LaSb and LaBi is that LaBi has band inversions at the time-reversal-invariant X points, which makes it topologically nontrivial. Compared with Sb atom, Bi atom has stronger SOC effect and its 6$p$ orbital electrons are more extended. On the other hand, the pressure is an effective approach to tune the electronic properties of materials, and the valence electrons generally become more extended due to the band broadening under pressure. Thus LaSb may also have band inversions at X points under certain pressure. From our above calculations (Fig. 4), LaSb does undergo a topological phase transition with the increase of pressure. Very recently, the nontrivial berry phase has been observed by both Shubnikov-de Haas and de Haas-Van Alphen oscillation experiments for LaBi \cite{Singha}. This is to say, the topological phase transition of LaSb under pressure may be confirmed by observing the nontrivial berry phase in experiment.

Another important property of LaSb is its extreme magnetoresistance \cite{CavaNP}. In semimetal LaSb, the valence and conduction bands have a small overlap [Fig. 2(b)]. If Sb atom can completely obtain three valence electrons of La, LaSb would become an insulator. Thus if all electrons in the conduction band of LaSb are transferred to the valence band, the latter would be fully filled. In this sense, it is easy to understand that the electron-like and hole-like carrier concentrations are equal in LaSb, as evidenced by our previous calculations \cite{GuoPRB} and related experiments \cite{LouPRL}. Meanwhile, transport experiments have discovered that the carrier mobilities are very high in LaSb \cite{CavaNP}. Accordingly, within the semiclassical band model, the electron-hole  compensation and high carrier mobilities can naturally explain the XMR effect\cite{CavaN, LeiNJP, GuoPRB}.

\begin{table}[!t]
\caption{\label{tab:II} The parities of all filled states of LaSb in NaCl-type structure at eight time-reversal-invariant points in the Brillouin zone after the topological phase transition.}
\begin{center}
\begin{tabular*}{0.99\columnwidth}{@{\extracolsep{\fill}}cccccccccc}
\hline\hline
  &      $\Gamma$ & L & L & L & L & X & X & X & total  \\
\hline
  1 & - & - & - & - & - & - & - & - & + \\
  3 & - & - & - & - & - & - & - & - & + \\
  5 & - & - & - & - & - & - & - & - & +  \\
  7 & + & - & - & - & - & + & + & + & + \\
  9 & - & + & + & + & + & - & - & - & + \\
  11 & - & + & + & + & + & - & - & - & +  \\
  13 & - & + & + & + & + & + & + & + & - \\
  total & + & + & + & + & + & - & - & - & 1 \\
\hline\hline
\end{tabular*}
\end{center}
\end{table}

In addition to the electron-hole compensation, the topological protection has also been proposed to explain the XMR effect. Since many XMR materials are topologically nontrivial \cite{HuangPRX, LiangNM, CavaN, LuPRB, YanNP}, it is very important to explore the relationship of XMR effect and topologically property. For LaSb, with increasing pressure, our calculations indicate that the small overlap of valence and conduction bands still exists (Fig. 3), thus the charge compensation also holds. Considering that the charge compensation accompanies with the the topological phase transition in LaSb under pressure, as shown by the calculations, thus if the MR of LaSb has sharp even discontinuous variation around the transition pressure in magnetic transport experiments, the topological property will have a substantial impact on the XMR, otherwise it will have no dominating effect on the XMR. In this sense, the NaCl-type LaSb under pressure is a promising platform to clarify this relationship.

\section{SUMMARY}
\label{sec:summary}

By using first-principles electronic structure calculations, we predict that the XMR material LaSb takes a topological phase transition without breaking any symmetry under hydrostatic pressure. Irrespective of topological phase transition, the electron-type and hole-type carriers retain compensated. Considering that both charge-compensation mechanism and topological protection mechanism have been proposed to explain the XMR effect, the pressed LaSb provides an ideal playground for studying the individual role of topological property playing in the XMR phenomenon.

\begin{acknowledgments}

We thank Yuan-Yao He, Tao Li, He-Chang Lei, and Shan-Cai Wang for helpful conversations. This work was supported by the National Natural Science Foundation of China (Grants No. 11474356 and No. 91421304), the Fundamental Research Funds for the Central Universities, and the Research Funds of Renmin University of China (Grants No. 14XNLQ03 and No. 16XNLQ01). Computational resources have been provided by the Physical Laboratory of High Performance Computing at Renmin University of China.

\end{acknowledgments}


\begin{thebibliography}{}

\bibitem{Hasan} M. Z. Hasan and C. L. Kane, Rev. Mod. Phys. \textbf{82}, 3045 (2010).
\bibitem{Qi} X.-L. Qi and S.-C. Zhang, Rev. Mod. Phys. \textbf{83}, 1057 (2011).
\bibitem{Hong} H.-M. Weng, X. Dai, and Z. Fang, J. Phys.: Condens. Matter \textbf{28}, 303001 (2016).
\bibitem{Thouless} D. J. Thouless, M. Kohmoto, M. P. Nightingale, and M. den Nijs, Phys. Rev. Lett. \textbf{49}, 405 (1982).
\bibitem{Wen}X.-G. Wen, Adv. Phys. \textbf{44}, 405 (1995).
\bibitem{HongNa3Bi}Z.-J. Wang, Y. Sun, X.-Q. Chen, C. Franchini, G. Xu, H.-M. Weng, X. Dai, and Z. Fang, Phys. Rev. B \textbf{85}, 195320 (2012).
\bibitem{ChenS} Z.-K. Liu, B. Zhou, Y. Zhang, Z.-J. Wang, H.-M. Weng, D. Prabhakaran, S.-K. Mo, Z.-X. Shen, Z. Fang, X. Dai, Z. Hussain, Y.-L. Chen, Science \textbf{343}, 864 (2014).
\bibitem{HongCdAs}Z.-J. Wang, H.-M. Weng, Q. Wu, X. Dai, and Z. Fang, Phys. Rev. B \textbf{88}, 125427 (2013).
\bibitem{ChenN} Z.-K Liu, $et$ $al$.  Nat. Mater. \textbf{13}, 677 (2014).
\bibitem{HongPRX} H.-M Weng, C. Fang, Z. Fang, B. A. Bernevig, and X. Dai, Phys. Rev. X \textbf{5}, 011029 (2015).
\bibitem{DingPRX} B. Q. Lv, H. M. Weng, B. B. Fu, X. P. Wang, H. Miao, J. Ma, P. Richard, X. C. Huang, L. X. Zhao, G. F. Chen, Z. Fang, X. Dai, T. Qian, and H. Ding, Phys. Rev. X \textbf{5}, 031013 (2015).
\bibitem{HasanS} S.-Y. Xu, I. Belopolski, N. Alidoust, M. Neupane, G. Bian, C.-L. Zhang, R. Sankar, G.-Q. Chang, Z.-J. Yuan, C.-C. Lee, S.-M. Huang, H. Zheng, J. Ma, D. S. Sanchez, B.-K. Wang. A. Bansil, F.-C. Chou, P. P. Shibayev, H. Lin, S. Jia, and M. Z. Hasan, Science \textbf{349}, 613 (2015).
\bibitem{DingNP} B. Q. Lv, N. Xu, H. M. Weng, J. Z. Ma, P. Richard, X. C. Huang, L. X. Zhao, G. F. Chen, C. E. Matt, F. Bisti, V. N. Strocov, J. Mesot, Z. Fang, X. Dai, T. Qian, M. Shi, and H. Ding, Nat. Phys. \textbf{11}, 724 (2015).
\bibitem{Abrikosov} A. A. Abrikosov, Phys. Rev. B \textbf{58}, 2788 (1998).
\bibitem{Neupane} M. Neupane, S.-Y. Xu, R. Sankar, N. Alidoust, G. Bian, C. Liu, I. Belopolski, T.-R. Chang, H.-T. Jeng, H. Lin, A. Bansil, F.-C. Chou and M. Z. Hasan, Nature Commun. \textbf{5}, 3786 (2014).
\bibitem{LiangNM} T. Liang, Q. Gibson, M. N. Ali, M. Liu, R. J. Cava, and N. P. Ong, Nat. Mater. \textbf{14}, 280 (2015).
\bibitem{YanNP}C. Shekhar, A. K. Nayak, Y. Sun, M. Schmidt, M. Nicklas, I. Leermakers, U. Zeitler, Z.-K. Liu, Y.-L. Chen, W. Schnelle, J. Grin, C. Felser, and B.-H. Yan, Nat. Phys. \textbf{11}, 645 (2015).
\bibitem{Nielsen} H. B. Nielsen and M. Ninomiya, Phys. Lett. B \textbf{130}, 389 (1983).
\bibitem{VishwanathPRX}S. A. Parameswaran, T. Grover, D. A. Abanin, D. A. Pesin, and A. Vishwanath, Phys. Rev. X \textbf{4}, 031035 (2014).
\bibitem{XiongS} J. Xiong, S. K. Kushwaha, T. Liang, J. W. Krizan, M. Hirschberger, W.-D. Wang, R. J. Cava, N. P. Ong, Science \textbf{350}, 413 (2015).
\bibitem{HuangPRX} X.-C. Huang, L.-X. Zhao, Y.-J. Long, P.-P.Wang, D. Chen, Z.-H. Yang, H. Liang, M.-Q. Xue, H.-M. Weng, Z. Fang, X. Dai, and G.-F. Chen, Phys. Rev. X \textbf{5}, 031023 (2015).
\bibitem{CavaNP} F. F. Tafti, Q. D. Gibson, S. K. Kushwaha, N. Haldolaarachchige, and R. J. Cava, Nat. Phys. \textbf{12}, 272 (2016).
\bibitem{LouPRL}L.-K. Zeng, R. Lou, D.-S. Wu, Q. N. Xu, P.-J. Guo, L.-Y. Kong, Y.-G. Zhong, J.-Z. Ma, B.-B. Fu, P. Richard, P. Wang, G.-T. Liu, L. Lu, Y.-B. Huang, C. Fang, S.-S. Sun, Q. Wang, L. Wang, Y.-G. Shi, H.-M. Weng, H.-C. Lei, K. Liu, S.-C. Wang, T. Qian, J.-L. Luo,and H. Ding, Phys. Rev. Lett. \textbf{117}, 127204 (2016).
\bibitem{LeiNJP} S.-S. Sun, Q. Wang, P.-J. Guo, K. Liu, and H.-C. Lei, New J. Phys. \textbf{18}, 082002 (2016).
\bibitem{CavaN} M. N. Ali, J. Xiong, S. Flynn, J. Tao, Q. D. Gibson, L. M. Schoop, T. Liang, N. Haldolaarachchige, M. Hirschberger, N. P. Ong, and R. J. Cava, Nature (London) \textbf{514}, 205 (2014).
\bibitem{FengPRL} J. Jiang, F. Tang, X.-C. Pan, H.-M. Liu, X.-H. Niu, Y.-X. Wang, D.-F. Xu, H.-F. Yang, B.-P. Xie, F.-Q. Song, P. Dudin, T. K. Kim, M. Hoesch, P. K. Das, I. Vobornik, X.-G. Wan, and D.-L. Feng, Phys. Rev. Lett. \textbf{115}, 166601 (2015).
\bibitem{ShenPRL}J.-F. He, C.-F. Zhang, N. J. Ghimire, T. Liang, C.-J. Jia, J. Jiang, S-J. Tang, S.-D. Chen, Y. He, S.-K. Mo, C. C. Hwang, M. Hashimoto, D.-H. Lu, B. Moritz, T. P. Devereaux, Y.-L. Chen, J.-F. Mitchell, and Z.-X. Shen, Phys. Rev. Lett. \textbf{117}, 267201 (2016).
\bibitem{GuoPRB} P.-J. Guo, H.-C. Yang, B.-J. Zhang, K. Liu, and Z.-Y. Lu, Phys. Rev. B \textbf{93}, 235142 (2016).
\bibitem{FengPRB} X.-H. Niu, D.-F. Xu, Y.-H. Bai, Q. Song, X.-P. Shen, B.-P. Xie, Z. Sun, Y.-B. Huang, D. C. Peets, and D.-L. Feng, Phys. Rev. B \textbf{94}, 165163 (2016).
\bibitem{GaoPRB} L. Gao, Y.-Y. Xue, F. Chen, Q. Xiong, R.-L. Meng, D. Ramirez, C.-W. Chu, J. H. Eggert, and H.-K. Mao, Phys. Rev. B \textbf{50}, 4260(R) (1994).
\bibitem{ChiPRL}Z.-H. Chi, X.-M. Zhao, H.-D. Zhang, A. F. Goncharov, S. S. Lobanov, T. Kagayama, M. Sakata, and X.-J. Chen, Phys. Rev. Lett. \textbf{113}, 036802 (2014).
\bibitem{Blochl}P. E. Bl$\ddot{\text{o}}$chl, Phys. Rev. B \textbf{50}, 17953 (1994).
\bibitem{Kresse1999}G. Kresse and D. Joubert, Phys. Rev. B \textbf{59}, 1758 (1999).
\bibitem{Kresse1993} G. Kresse and J. Hafner, Phys. Rev. B \textbf{47}, 558 (1993).
\bibitem{KresseCMS} G. Kresse and J. Furthm$\ddot{\text{u}}$ller, Comput. Mater. Sci. \textbf{6}, 15 (1996).
\bibitem{Kresse1996} G. Kresse and J. Furthm$\ddot{\text{u}}$ller, Phys. Rev. B \textbf{54}, 11169 (1996).
\bibitem{PerdewJPC} J. P. Perdew, A. Ruzsinszky, J.-M. Tao, V. N. Staroverov, G. E. Scuseria, and G. I. Csonka, J. Chem. Phys. \textbf{123}, 062201 (2005).
\bibitem{PerdewPRL} J. P. Perdew, K. Burke, and M. Ernzerhof, Phys. Rev. Lett. \textbf{77}, 3865 (1996).
\bibitem{hybrid} J. Heyd, G. E. Scuseria, and M. Ernzerhof, J. Chem. Phys. \textbf{118}, 8207 (2003); J. Chem. Phys. \textbf{124}, 219906 (2006).
\bibitem{hse06} A. V. Krukau, O. A. Vydrov, A. F. Izmaylov, and G. E. Scuseria, J. Chem. Phys. \textbf{125}, 224106 (2006).
\bibitem{Ye13PRB} Q.-Q. Ye, K. Liu, and Z.-Y. Lu, Phys. Rev. B \textbf{88}, 205130 (2013).
\bibitem{Fu07PRB} L. Fu and C. L. Kane, Phys. Rev. B \textbf{76}, 045302 (2007).
\bibitem{Leger} J. M. Leger, D. Ravot, and J. Rossat-Mignod, J. Phys. C: Solid State Phys. \textbf{17}, 4935 (1984).
\bibitem{Becke06} A. D. Becke and E. R. Johnson, J. Chem. Phys. \textbf{124}, 221101 (2006).
\bibitem{Tran09} F. Tran and P. Blaha, Phys. Rev. Lett. \textbf{102}, 226401 (2009).
\bibitem{FuPRL} L. Fu, C. L. Kane, and E. J. Mele, Phys. Rev. Lett. \textbf{98}, 106803 (2007).
\bibitem{NayakNC} J. Nayak, S.-C. Wu, N. Kumar, C. Shekhar, S. Singh, J. Fink, E. E. D. Rienks, G. H. Fecher, S. S. P. Parkin, B. Yan, and C. Felser, Nat. Commun. \textbf{8}, 13942 (2017).
\bibitem{LouPRB}R. Lou, $et$ $al$. Phys. Rev. B \textbf{95}, 115140 (2017).
\bibitem{Singha}R. Singha, B. Satpati, and P. Mandal, arXiv:1703.06100.
\bibitem{LuPRB}J. Feng, Y. Pang, D. Wu, Z. Wang, H.-M Weng, J. Li, X. Dai, Z. Fang, Y.-G Shi, and L. Lu, Phys. Rev. B \textbf{92}, 081306 (2015).

\end{thebibliography}
\end{document}